\documentclass[sigconf]{acmart}

\def\BibTeX{{\rm B\kern-.05em{\sc i\kern-.025em b}\kern-.08emT\kern-.1667em\lower.7ex\hbox{E}\kern-.125emX}}
    
\copyrightyear{2019} 
\acmYear{2019} 
\acmConference[RecSys '19]{Thirteenth ACM Conference on Recommender Systems}{September 16--20, 2019}{Copenhagen, Denmark}
\acmBooktitle{Thirteenth ACM Conference on Recommender Systems (RecSys '19), September 16--20, 2019, Copenhagen, Denmark}
\acmPrice{15.00}
\acmDOI{10.1145/3298689.3347029}
\acmISBN{978-1-4503-6243-6/19/09}

\setcopyright{acmcopyright}
\begin{document}

%
\title[Music Recommendations in Hyperbolic Space]{Music Recommendations in Hyperbolic Space: An Application of Empirical Bayes and Hierarchical Poincar\'e Embeddings}

%


\author{Timothy Schmeier}
\email{Timothy.Schmeier@pepsico.com}
\affiliation{%
  \institution{iHeartRadio}
  \city{New York City}
  \state{New York}
}
%

\author{Sam Garrett}
\email{sam@iheart.com}
\affiliation{%
  \institution{iHeartRadio}
  \city{New York City}
  \state{New York}
}

\author{Joseph Chisari}
\email{jchisari@iheart.com}
\affiliation{%
  \institution{iHeartRadio}
  \city{New York City}
  \state{New York}
}

\author{Brett Vintch}
\email{brett@iheart.com}
\affiliation{%
  \institution{iHeartRadio}
  \city{New York City}
  \state{New York}
}

%

%
\begin{abstract}
Matrix Factorization (MF) is a common method for generating recommendations, where the proximity of entities like users or items in the embedded space indicates their similarity to one another. Though almost all applications implicitly use a Euclidean embedding space to represent two entity types, recent work has suggested that a hyperbolic Poincar\'e ball may be more well suited to representing multiple entity types, and in particular, hierarchies. We describe a novel method to embed a hierarchy of related music entities in hyperbolic space. We also describe how a parametric empirical Bayes approach can be used to estimate link reliability between entities in the hierarchy. Applying these methods together to build personalized playlists for users in a digital music service yielded a large and statistically significant increase in performance during an A/B test, as compared to the Euclidean model.
\end{abstract}

%
%

\ccsdesc[500]{Information systems~Recommender systems}
\ccsdesc[500]{Computing methodologies~Machine learning}

%
\keywords{recommender system; poincare; hyperbolic; hierarchical}

%
\maketitle

\section{Introduction}
Recommendation systems are an important tool for user personalization in most modern digital media services. Entities are commonly embedded in a high dimensional Euclidean space with Matrix Factorization (MF), and recommendations for a given user (or item) are made as a function of proximity to other entities \cite{bobadilla2013recommender, johnson2014logistic}. As a streaming digital music service, 
iHeartRadio employs a number of such linear recommendation systems across multiple entity types to personalize the user experience and assist discovery.

Despite the wide use and success of MF models, an inherent limitation is the linearity of Euclidean space. This is especially true for models that must learn to embed more than two types of entities. While this situation is frequently encountered with the inclusion of user or item side information, it is also encountered when items have directional relationships as in a hierarchy. Previous research has outlined strategies for incorporating linear pairwise interactions between entities, such as Collective Matrix Factorization \cite{singh2008relational} and Factorization Machines \cite{rendle2010factorization, juan2016field}. While these methods allow for the inclusion of multiple entity types, they do not assist in representing directed hierarchies.

Nickel \& Kiela introduced the use of a Poincar\'e ball for embedding known hierarchical networks in a high dimensional hyperbolic space \cite{nickel2017Poincare}. Later, Vinh \textit{et al.} showed that this approach naturally extends to the collaborative filtering task of decomposing a user-item interaction matrix. In this case, the embeddings live in a Poincar\'e ball as opposed to a Euclidean inner product space \cite{vinh2018hyperbolic}. The authors posit that the curvature of hyperbolic space allows for the embedding of these simple two-entity hierarchical networks with fewer parameters and higher contrast. Inspired by this work, we hypothesize that Poincar\'e ball embeddings of the deeper and richer hierarchical network of musical entities, such as artists, radio stations, and genres may produce a powerful item-based recommendation system.

Directed links between entities is a core element of the Poincar\'e embedding formulation. In the first demonstrations, embeddings were fit to axiomatic links from known taxonomies. For example, Nickel \textit{et al.} show embeddings for a taxonomy of mammals. For music recommendations some directed links fit this pattern well, such as an artist belonging to a genre or a radio station belonging to a programming format. However, other directed relationships must be learned from user behavior. We introduce the use of a parametric empirical Bayes-based lower credibility interval to discover links between entities.

Our primary contribution describes a novel and flexible method for creating a hierarchical network based on parametric empirical Bayes techniques. This approach accommodates any data that can be modeled as conjugate distributions, unifies data sets resulting from different generating processes, and leverages item side information into a single coherent network. We show that the embedding of this network into a Poincar\'e  ball generates a recommendation system capable of outperforming comparable methods in an A/B test.

\section{Methods} \label{methods_section}

\subsection{Strategy}
iHeartRadio is a digital music service that, among other offerings, gives users access to two radio-like music products: live radio stations, which are simulcast from traditional broadcast stations, and custom artist stations, which are coherent radio stations that are seeded by the selection of an artist. In this work we embed five types of entities derived from these products into a Poincar\'e ball. The overall flow of section \ref{methods_section} proceeds as follows. First, we describe how each entity type relates to one another in a hierarchical graph. Next, we introduce a parametric empirical Bayes approach to determine which specific entities in the graph have a strong enough link to include in the model. Finally, we describe how links are embedded in a Poincar\'e space, and how this model is used to generate recommendations for users.

\subsection{Hierarchical entity graph}

\begin{figure}[t!]
\begin{center}
\includegraphics[width=\linewidth]{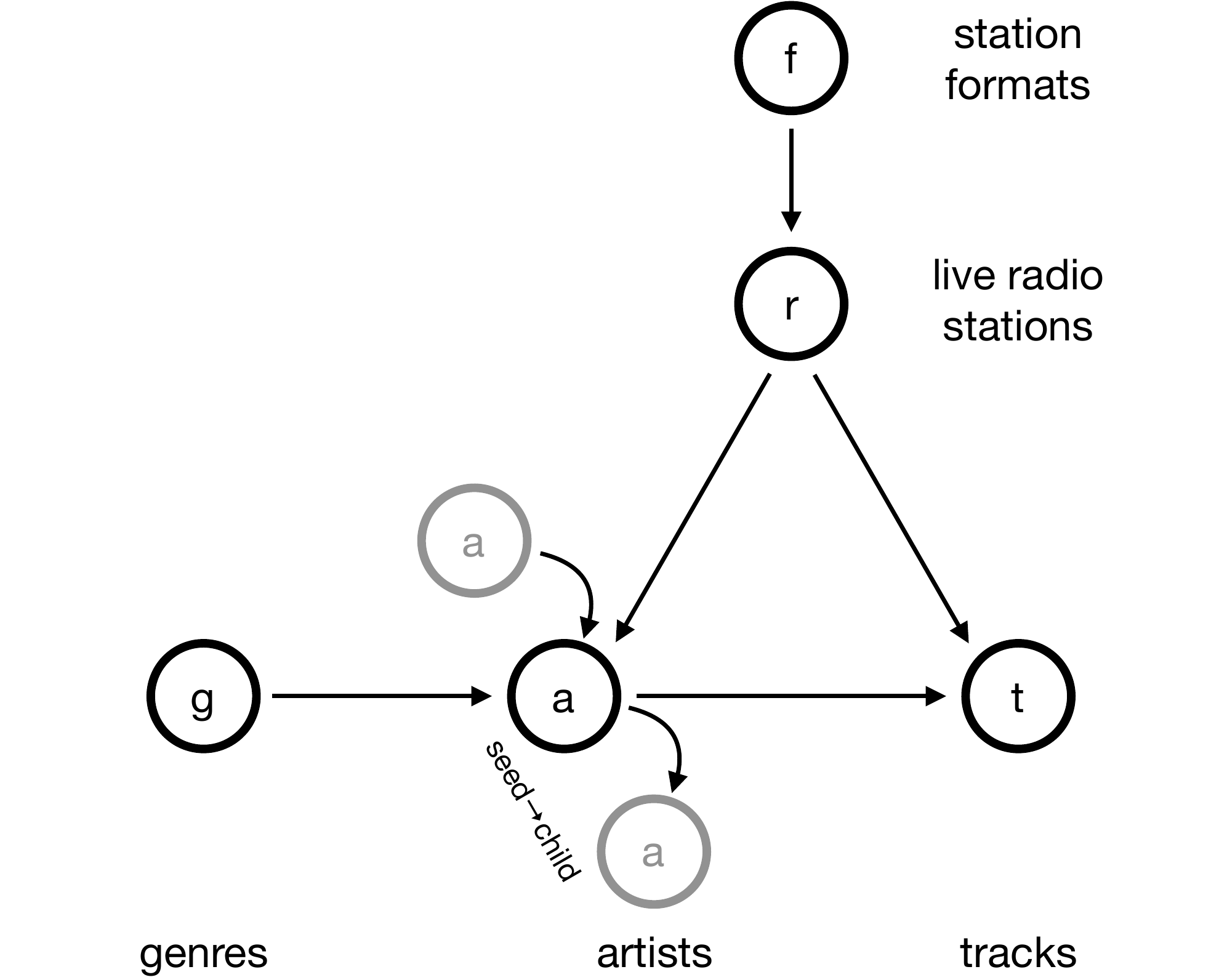} 
\end{center}
\caption[network]{A graph relating five entity types captures logical, directed pairwise relationships between each type of entity. Note that custom artist stations do not appear in this graph, as all links are attributed to the seed artist from which they were created.}
\label{fig:network}
\end{figure}

The five types of entities that we model are live radio stations, live radio station format, artists, artist genre, and tracks. Poincar\'e embeddings are designed to represent hierarchical relationships between entities, and our five entity types form natural pairwise hierarchies; stations play artists and tracks while the stations themselves belong to a high-level radio format description, and artists can be additionally described with a genre label. We also allow for directional links \textit{between} artists to capture relationships that arise from custom artist stations; that is, tracks in a station by artists other than the station's seeded artist are deemed child artists for the station seed artist. Though this last type of pairwise link is not strictly hierarchical (two seed artists can also be child artists in each other's stations, creating some cyclical links), experimentation indicated that this was an important signal for our model. The hierarchical network is depicted in Figure \ref{fig:network}. Note that links between entity types are many to many.

\subsection{Data} \label{data_section}

iHeartRadio is both a digital and analog music provider; we broadcast tens of thousands of tracks over hundreds of analog stations every month. To construct our data set we collect all track spins over all stations over a lookback window of 6 months. We consider a track's play rate (the spin count over the number of days presented) in a station to be a raw measure of link strength between the track and station. Both the spin counts and number of days over which a track is presented vary over orders of magnitude. 

In addition to simulcasting analog stations, we allow users the ability to create custom artist station that are seeded with an artist of their choice. We pull child track counts for each seed artist station over the same lookback window of 6 months. The measure of raw link strength for tracks in a seed artist station is the number of completions in the station over all starts (user skips and station exits count against track completion). The raw measure of link strength for child artists in a seed artists station is the summation over their track's link strengths that are present in the seed artist station. Seed artist station starts, and track counts within each station, also vary over a large order of magnitude.

To complete our graph, we leverage internal dimensional data that describes the format of each live radio station (e.g. Urban or Contemporary Hits Radio (CHR)) and the genre of each artist (e.g. Classic Rock or Country). The order of magnitude of each entity type count can be found in Table 1.


\begin{table}[h!]
  \begin{center}
    \caption{Entity counts}
    \label{tab:counts}
    \begin{tabular}{l|r|r} 
      Entity type & Order of magnitude \\
      \hline
      Station formats and genres & 10-100 \\
      Live radio stations & 1,000 \\
      Artists & 10,000  \\
      Tracks & 1,000,000 \\
      Users & 1,000,000\\
    \end{tabular}
  \end{center}
  \label{tab:counts}
\end{table}

\subsection{Empirical Bayes estimation of link quality}

The embedding model described in section \ref{poincare_section} has no notion of link weight. Thus, it is imperative that we set criteria for those links that are strong enough to be included in the model and those which should be ignored. For example, a track that plays in a live station only once every few months should not be given the same prominence as a track that plays every hour. To this end, we describe a parametric empirical Bayes method to estimate link strength, and confidence thresholds to determine which links should be included and which should be ignored.

Let $A$ be the set of artists, each of whom have an associated digital artist station $S$. Let $T_s \subset T$ be the set of tracks which are presented to users within the context of digital artist station $s$. Across actions from all users within the artist station $s$, each track has an observed number of trials (presentations to users), $n_{st}$, and number of successes (completions), $k_{st}$. The likelihood for the completion rate, or probability $p_{st}$, is modeled as a binomial distribution:

\begin{displaymath}
k_{st}\ |\ s \sim Binomial(n_{st}, \ p_{st})\ \forall t \in T_{s}.
\end{displaymath}

Similarly, let $T_r \subset T$ be the set of tracks which have been spun on broadcast radio station $r$, where $R$ is the set of broadcast radio stations. Each element $t$ in $T_r$ has a count of radio spins $n_{rt}$ in broadcast station $r$ in a distinct number of days in the lookback window $d$. The likelihood of track spin counts for r can be modeled as a Poisson with rate parameter $\lambda_{rt}$:

\begin{displaymath}
n_{rt }\ / \ d\ |\ r\  \sim Poisson(\lambda_{rt})\ \forall t \in T_{r}.
\end{displaymath}

Next, we choose conjugate priors to these two likelihood functions (sampling techniques like Markov Chain Monte Carlo proved intractable due to the scale of our data). The conjugate prior of the binomial likelihood is the beta distribution while the conjugate prior of the Poisson likelihood is a gamma distribution:

\begin{gather*}
p_{s}\ |\ s \sim Beta(\alpha_{s},\ \beta_{s}) \\
\lambda_{\text{r\ }}|\ r \sim Gamma(\alpha_{r},\ \beta_{r})
\end{gather*}

We use observed completion data for the tracks $T_s$, and play data for $T_r$ to solve for the parameters of these two prior distributions with maximum likelihood estimation (MLE). This method of using observed data to solve for the prior, followed by an update rule to obtain a posterior distribution, is known as empirical Bayes Estimation \cite{casella1985introduction} (this is in contrast to sampling techniques, which solve all distributions simultaneously).

The posterior beta and gamma distributions for the beta-binomial and Poisson-Gamma models respectively, are obtained with the following update rules:

\begin{gather*}
p_{st}\ |\ s,\ k_{st},\ n_{st} \sim Beta(\alpha_{s} + k_{st},\ \beta_{s} + \ n_{st} - k_{st})\ \forall t \in T_{s} \\
\lambda_{rt}\ |\ r,\ n_{rt},\ d\  \sim Gamma(\alpha_{r}\  + \ n_{rt},\ \beta_{r}\  + \ d)\ \forall t \in T_{r}.
\end{gather*}

As we note in section \ref{data_section}, track occurrences, $n_{st}$ and $n_{rt}$, vary over several orders of magnitude within both custom artist and live radio stations. For custom artist stations, MLE point estimates of success rates, $p_{st} = k_{st}\ /\ n_{st}$, can have large variance when the small number of trials, $n_{st}$, is small. However, the Bayesian posterior distribution point estimates $\mu_{st} = {(\alpha}_{s} + k_{st})\ /\ (\beta_{s} + \ n_{st} - k_{st})$ exhibit a feature known as shrinkage. This effect imparts the model with a sense of skepticism for estimates that deviate strongly from the prior distribution by pulling them closer to the prior. We observe that this a desirable property for our model in that it moderates extreme estimates for tracks with few trials. A similar issue exists with rate estimates for tracks with few observed days $d$ in the broadcast station data set. Again, the effect of the Bayesian model is a desirable moderation of extreme rate estimates over short time periods.

\subsection{Link creation and Poincar\'e embeddings} \label{poincare_section}

We define a track score as the the lower bound of the posterior distribution credibility interval. For custom artist radio, $\text{score}_{st} = P_{st}(\alpha_{} = 0.05)$. This definition ensures that distributions with large variance are penalized. That is, a large accumulation of positive evidence is required for a track to achieve high score within a station.

Directed links $\{ a \rightarrow t\ |\ t \in T_{s},\ \text{score}_{st}\  > \ P_{75}\}$ are created from artist $a$ to the top quartile of tracks in their station $s$. In this way, each artist is defined by the best performing tracks from their station. Note that many artists can make links to the same track. Though not strictly hierarchical, we also create directed links between artists. For the tracks in $T_{s}$ there exist a proper subset of tracks by 'child' artists $B_{s} \subset A$, which are all other artists appearing in the station that are not the seeded artist. Summing $\text{score}_{st}$ over the tracks belonging to the child artists in $B$ provides a child artist score:

\begin{displaymath}
{\text{score}_{\text{sb}} = \sum_{t} \text{score}_{\text{stb}}\ |\ b \in}_{}B_{s}
\end{displaymath}

Summation was chosen as opposed to other aggregation functions because summation naturally takes into account both the cardinality and the quality of the set of tracks from a child artist within the context of the parent digital artists station. Directed links $\{ a \rightarrow b\ |\ b \in B_{s},\ \text{score}_{sb}\  > \ P_{75}\}$ are created from artists $a$ to each child artist $b$ in the top quartile. We also create directed links between an artist's primary genre label and the artist; where $G$ is the set of genres that covers all artists in our data set, $\{ g \rightarrow a\}$.

Similarly, for live radio stations we create a score using the value at the lower credible interval of the gamma posterior distribution, ${\text{score}_{rt} = P}_{rt}(\alpha_{} = 0.05)$ to create directed links from the live station to the best performing child tracks, $\{ r \rightarrow t\ |\ t \in T_{r},\ \text{score}_{rt}\  > \ P_{75}\}$. Summing over the tracks that have been authored by child artists $B_r$ provides a child artist score:

\begin{displaymath}
	{\text{score}_{rb} = \sum_{t} \text{score}_{rb}\ |\ b \in}_{}B_{r}
\end{displaymath}

We create directed links between each live station and their best performing child artists $\{ r \rightarrow b\ |\ b \in B_{r},\ \text{score}_{\text{rb}}\  > \ P_{75}\}$. Each live station is associated with a programming format, $f$, which is also used to form links $\{ f \rightarrow r\}$ for relevant tuples.

\begin{figure}[t!]
\begin{center}
\includegraphics[width=\linewidth]{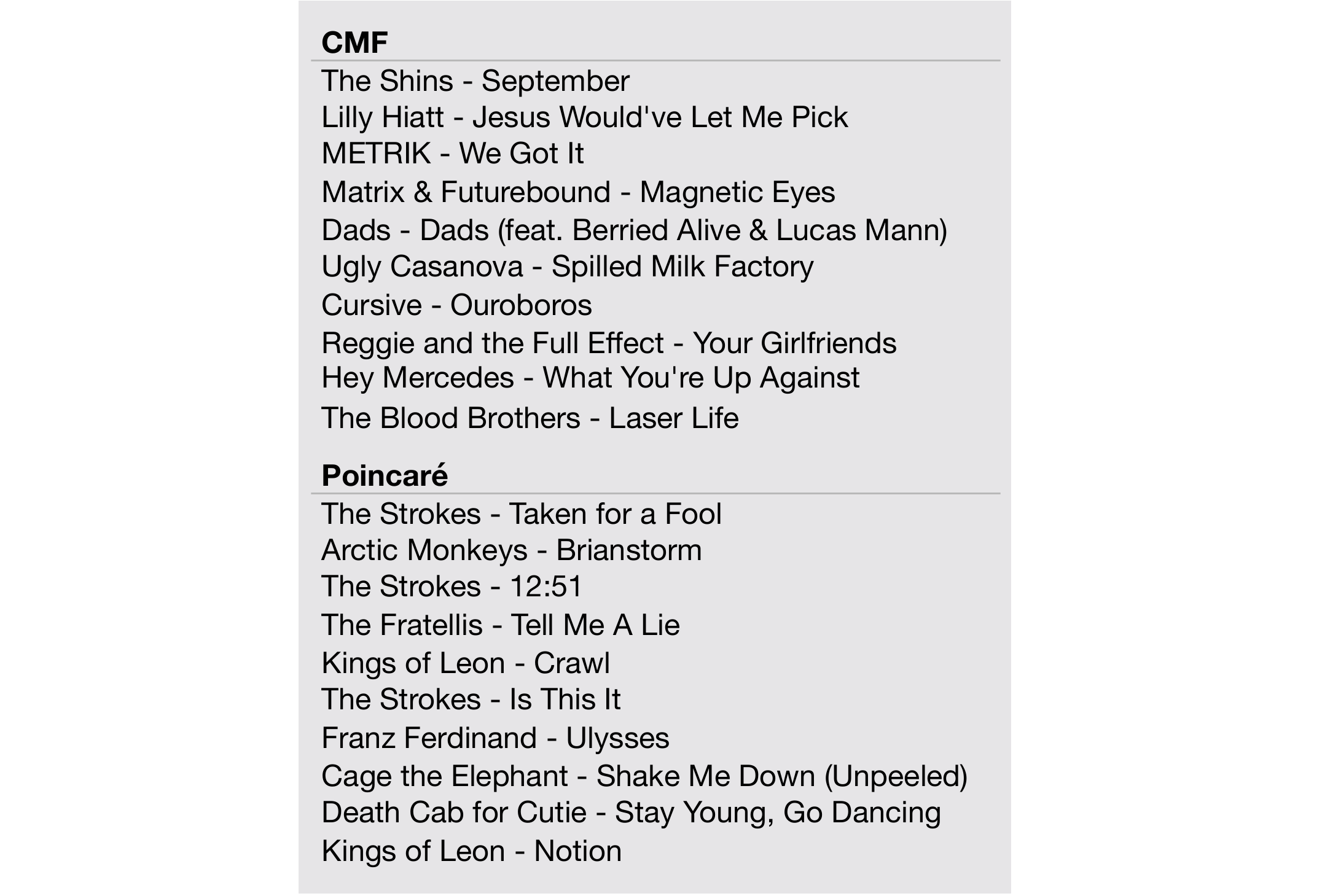} 
\end{center}
\caption[A2A]{Track recommendations for Modest Mouse from the CMF model and the Poincar\'e model.}
\label{fig:A2A}
\end{figure}

A training data set is constructed as the set of tuples containing all directed links described above. All together, there are 6 types of links between 5 distinct entity groups. We prune any entity with fewer than 20 total links. This data set was used to fit embeddings for each entity in hyperbolic Poincar\'e space. We used the Poincar\'e model implementation from the gensim open source software package \cite{rehurek2010software} to embed a Poincar\'e ball of rank 15.

\subsection{Personalized playlist experiment} \label{section_experiments}

iHeartRadio has a personalized playlist feature called \textit{Weekly Mixtape}. This feature is powered by a combination of recent user listening history, business rules, and a variant of Collective Matrix Factorization (CMF) between artist genre, and track entities. The CMF algorithm factorizes one ratings matrix into rank 100 vectors, where all user-entity matrices are concatenated together through the user. For one week we ran an A/B test experiment where 70\% of users received the default (control) experience, and 30\% of users received an experience where the CMF algorithm was swapped out in favor of the Poincar\'e model. Recommendations for both models are generated by finding content nearby a user's recent listening history, subject to business rules. We use hyperbolic distance for the Poincar\'e model and Cosine distance for the CMF model to search for nearby content. We measured the average time spent listening to the personalized playlists for users in both groups.

\section{Results}

We begin by examining the relative quality of recommendations from two models: the Poincar\'e model and a Collective Matrix factorization model (both described in section \ref{methods_section}). For popular artists (artists in the top 10\textsuperscript{th} percentile of total listening), we observe that the two models largely agree on tasks like recommending child artists for a seed artist, and recommending tracks for a seed artist. Model qualities seem to diverge for less popular artists, particularly for hierarchical tasks like recommending tracks for artists. We show examples of this phenomenon for the artist Modest Mouse (Figure \ref{fig:A2A}).

\begin{figure}[t!]
\begin{center}
\includegraphics[width=\linewidth]{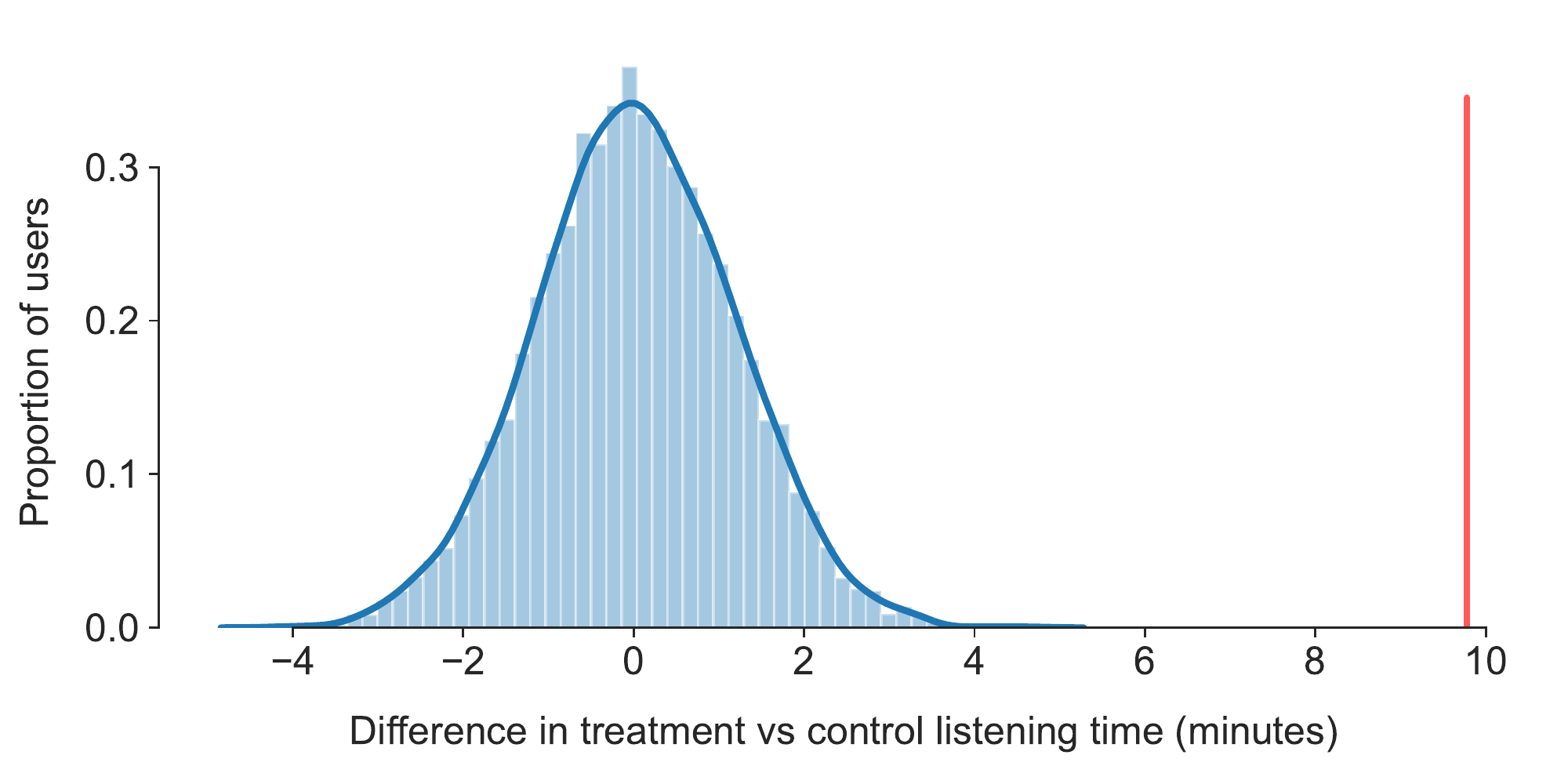} 
\end{center}
\caption[bootstrap]{A/B test group assignment is bootstrapped 10,000 times, with assignments randomized each run. The distribution of segment differences for these bootstrapped runs is shown in blue (along with a kde fit), while actual observed difference is shown in red. No bootstrapped run occurs to the right of the red line.}
\label{fig:bootstrap}
\end{figure}

\begin{figure}[t!]
\begin{center}
\includegraphics[width=\linewidth]{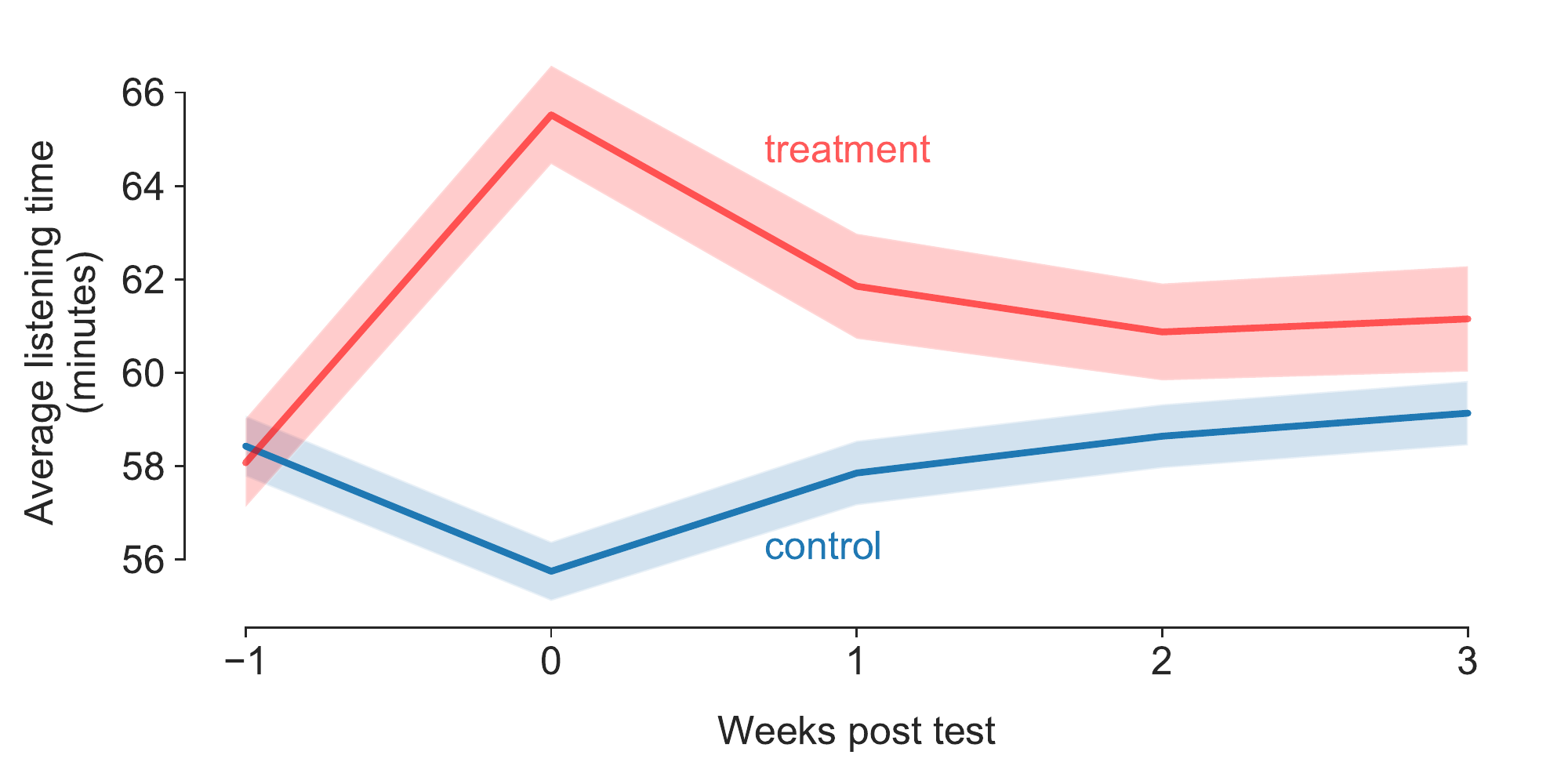} 
\end{center}
\caption[halo]{Differences between the test and treatment groups continue past the week of the test; this shows an increased likelihood for users to return to the product. Bands are 95\% confidence intervals.}
\label{fig:halo}
\end{figure}

We further test recommendation quality by incorporating them in iHeartRadio's personalized playlist product, \textit{Weekly Mixtape}. Treated users receive recommendations from the Poincar\'e model while control users receive recommendations from CMF (see section \ref{section_experiments}). Average Listening Time (ALT) for the treated group was 17.4\%, or 9.72 minutes, greater than the control group during the test period, which was significant under a bootstrap permutation test (Figure \ref{fig:bootstrap}; 10,000 permutations, p<0.0001). Furthermore, users exposed to the test condition continued to display increased ALT for up to three weeks after the conclusion of the A/B test, demonstrating the long lasting impact of the treatment (Figure \ref{fig:halo}). We note that it is unlikely that these results are due purely to a user novelty effect  \cite{kohavi2012trustworthy}; for other experiments with our user base we typically find that our users exhibit a primacy effect rather than a novelty effect. The performance of the treatment group was large enough to justify deploying the Poincar\'e model as the default playlist generator at the time of publication.

\section{Discussion and Future Work}

We have introduced the use of hyperbolic, or Poincar\'e, embeddings to represent hierarchical entities in a digital music service. We also introduced the use of empirical Bayes methodology to reliably estimate links between entities in real data sets. The resulting model is able to generate high quality recommendations that we judge to be qualitatively superior to Collective Matrix Factorization. Moreover, when tested in a personalized playlist setting, the Poincar\'e embeddings performed significantly better with real users, and the difference was so large that it persisted beyond the conclusion of the experiment.

Creating a hierarchical network using the described parametric empirical Bayes methods and embedding the resulting network in the Poincar\'e ball are both parallelizable operations resulting in an efficient, highly scalable algorithm \cite{nickel2017Poincare}. For the application of recommending music, we demonstrate that hyperbolic space produces a more useful embedding model than comparable matrix factorization techniques, and does so with fewer parameters. Moreover, the hierarchical structure of the data and embeddings means that high-level concepts like genre and station format do not need to live in the same subspace as low-level entities like tracks. This has implications for fitting the model, but also suggests new paradigms for music recommendation through the use of higher-level music concepts.

An important input into the Poincar\'e model is links between entities, and we introduce a novel way for estimating link reliability with empirical Bayes methods. These methods are flexible for data from range of likelihood and conjugate distributions and can be used to unify data sets from different generating processess.

Bayesian methods are also especially important for digital music providers, where the cold start problem for items is particularly acute; a single from a new artist might become a hit with ubiquitous airplay in a matter of days. Creating a new digital artist station or including their new hit singles into other relevant artist stations can be challenging with only sparse historical data. Some music platforms are known to inject these new artists and tracks into playlists or stations using manual curation. Others have employed musicians to analyze and tag relevant musical attributes in order to make content comparisons \cite{glaser2006consumer}. Our procedure helps to automatically embed new artists and tracks from broadcast and digital data with little or no human intervention. 

The methodology described here should naturally incorporate additional entity types beyond the five we choose to embed. Additionally we hypothesize that modifying the loss function of the embedding algorithm to accept continuous scores instead of discrete links, or incorporating triplet loss \cite{schroff2015facenet, vinh2018hyperbolic}, could allow the model to learn a more subtle embedding structure.

In conclusion, we have described a general method to construct a network hierarchy of related music entities from numerous data sets. We have described the use of parametric empirical Bayes in estimating link reliability in the construction of this network and shown that embedding this network in hyperbolic space produces a robust recommendation engine capable of outperforming traditional Matrix Factorization approaches in real world applications.

%

%
\bibliographystyle{ACM-Reference-Format}
\bibliography{bibliography}

\end{document}